# High-Resolution Modeling of Coronae and Winds in Solar-Type Stars with Varying Rotation Rates. I. X-Ray Coronae

Yue-Hong Chen (陈悦虹)[1,2,3], Julián D. Alvarado-Gómez[3], Xin Cheng (程鑫)[1,2], Yu Dai (戴煜)[1,2], Tong Shi[4], Katja Poppenhäger[3,5], Chen Xing (邢晨)[1,2], Shun Inoue[6], Jörn Warnecke[7,8], Maarit J. Korpi-Lagg[7,8,9], and Mingde Ding (丁明德)[2]

[1] School of Astronomy and Space Science, Nanjing University, Nanjing 210023, People's Republic of China; xincheng@nju.edu.cn
[2] Key Laboratory of Modern Astronomy and Astrophysics (Nanjing University), Ministry of Education, Nanjing 210023, People's Republic of China
[3] Leibniz Institute for Astrophysics Potsdam, Potsdam 14482, Germany
[4] SETI Institute, Mountain View, CA 94043, USA
[5] University of Potsdam, Institute of Physics and Astronomy, Potsdam-Golm 14476, Germany
[6] Department of Physics, Kyoto University, Kitashirakawa-Oiwake-cho, Sakyo-ku, Kyoto 606-8502, Japan
[7] Department of Computer Science, Aalto University, P.O. Box 15400, FI-00 076 Espoo, Finland
[8] Max-Planck-Institut für Sonnensystemforschung, Justus-von-Liebig-Weg 3, D-37077 Göttingen, Germany
[9] Nordita, KTH Royal Institute of Technology & Stockholm University, Hannes Alfvéns väg 12, SE-11419 Stockholm, Sweden
*Received 2025 September 5; revised 2025 October 8; accepted 2025 October 14; published 2025 December 8*

## Abstract

Stellar coronae are believed to be the main birthplace of various stellar magnetic activities. However, the structures and properties of stellar coronae remain poorly understood. Using the Space Weather Modeling Framework with the Alfvén Wave Solar Model (SWMF-AWSoM) and dynamo-generated surface magnetic maps, here we model the coronae of four solar-type stars. By incorporating the Sun, our work covers a range of stars with the rotation varying from 1.0 to $23.3\Omega_{\odot}$ (periods of 25–1 days). Guided by observations, we scale the magnetic field strength with increasing rotation, covering a range between 6.0 G and 1200 G approximately. In our models, energy release associated with small-scale magnetic flux is a key source of coronal heating and is essential for reproducing realistic coronal structures. Our models capture dense (1–2 orders of magnitude higher than solar values) and ultra-hot (∼10 MK) coronae dominated by closed field structures. Using the CHIANTI atomic database, we also compute synthetic X-ray spectra and derive the corresponding X-ray luminosities ($L_X$), which follow a scaling law to magnetic field $L_X \propto \langle |\boldsymbol{B}| \rangle^{1.75}$. Furthermore, the coronal X-ray emission is found to be rotationally modulated by the alternating presence of bright active regions and dark coronal holes. These results provide new insights into the extremely high-energy coronae of rapidly rotating solar-type stars, which differ markedly from the Sun.

*Materials only available in the online version of record: animations*

## 1. Introduction

The coronae of solar-type stars are characterized by a temperature of multimillion Kelvin, serving as the place of origin of high-energy electromagnetic radiation, notably X-rays (M. Güdel 2004; J. J. Drake & B. Stelzer 2023). They also generate continuous stellar winds (A. A. Vidotto 2021) and transient ejections of high-velocity, energetic particles (R. Schwenn 2006). These processes shape space weather and play a key role in determining the atmospheric retention and habitability of orbiting planets (I. Ribas et al. 2005; J. E. Owen & Y. Wu 2013; G. Gronoff et al. 2020). Understanding coronal structures and properties is thus essential for understanding both stellar evolution and its broader impact on planetary systems.

Coronal characteristics, particularly the density and temperature, are primarily governed by the stellar magnetic field, such as its strength, topology, and the filling factor. For example, observations of the O VII, Ne IX, and Mg XI, which form at the temperature of approximately 2 MK, show that low-activity stars typically exhibit low electron densities (e.g., $\log n_e < 10$, with $n_e$ in units of $cm^{-3}$), while densities in magnetically active stars might be $\log n_e =$ 10–13 (M. J. Aschwanden & A. O. Benz 1997; M. Güdel 2004; F. Reale 2007; M. Güdel & Y. Nazé 2009; R. Sasaki et al. 2021; S. Inoue et al. 2024). Nonetheless, cool and hot plasma components can reside in different magnetic structures or at different coronal heights, which requires multitemperature analysis. This is often achieved through spectroscopic diagnostics that allow the reconstruction of the emission measure distribution (EMD, $EM = n_e n_H V$ at a given temperature) or differential emission measure ($DEM = n_e n_H V / d \ln T$; e.g., J. Sanz-Forcada et al. 2019; M. Coffaro et al. 2020; J. Sanz-Forcada et al. 2025). Theoretically, DEM can be modeled based on coronal loop thermodynamics (R. Rosner et al. 1978; S. K. Antiochos 1980) and stochastic flare heating scenarios (M. Güdel et al. 2003). These models typically predict power-law relationships between DEM and temperature and are capable of reproducing certain observed properties. Observationally, stars with stronger X-ray fluxes ($F_X$) tend to exhibit larger total emission measures, regardless of spectral type (B. E. Wood 2018). However, the specific role of various stellar parameters in shaping the temperature and density structure of the corona remains not fully understood (M. Güdel & Y. Nazé 2009).

In low-mass stars, the X-ray luminosity ($L_X$) spans a wide range, from approximately $10^{25}$ to $10^{30}$ erg s$^{-1}$ in the 0.1–2.4 keV ROSAT bandpass, with the X-ray-to-bolometric luminosity ratio ($L_X/L_{bol}$) typically falling between $10^{-8}$ and $10^{-3}$ (J. J. Drake & B. Stelzer 2023).[10] For different stars, the $L_X$ exhibits a strong dependence on the stellar magnetic field. According to stellar dynamo theory (S. A. Barnes 2003; A. S. Brun & M. K. Browning 2017; P. Charbonneau & D. Sokoloff 2023), faster rotations usually lead to stronger magnetic fields. The stronger magnetic fields store and release more energy, leading to more efficient coronal heating and stronger X-ray emission. In rapidly rotating stars, $L_X/L_{bol}$ reaches a saturation regime, where the ratio becomes nearly constant and is no longer strongly dependent on rotation (R. Pallavicini et al. 1981; N. J. Wright et al. 2011; A. Reiners et al. 2022). For individual stars, $L_X$ can be modulated by rotational variability (S. Chandra et al. 2010; J. Sanz-Forcada et al. 2019) or long-term magnetic activity cycles (M. Coffaro et al. 2020; B. J. Wargelin et al. 2024), potentially driven by changes in the surface filling factors, active regions, and quiet regions.

Despite the observations described above, the coronal structures in solar-type stars cannot be resolved as they are for the Sun, due to a faraway distance.[11] This makes three-dimensional (3D) magnetohydrodynamic (MHD) modeling an effective and valuable tool to study the structures and physical properties of stellar coronae.

Generally, this type of coronal modeling critically depends on accurate photospheric conditions, particularly the surface magnetic field. Zeeman Doppler Imaging (ZDI) provides the essential boundary conditions (S. S. Vogt et al. 1987; M. Semel 1989; J. F. Donati et al. 1997), and has been widely used in 3D MHD simulations or models of stellar atmospheres (e.g., C. Garraffo et al. 2015; A. A. Vidotto et al. 2015; J. D. Alvarado-Gómez et al. 2016a, 2016b; J. D. do Nascimento et al. 2016; B. A. Nicholson et al. 2016; V. Réville et al. 2016; Q. Pognan et al. 2018; D. Ó Fionnagáin et al. 2019; D. Evensberget et al. 2021, 2022, 2023; J. J. Chebly et al. 2023). Nevertheless, one persistent challenge in ZDI-based modeling lies in flux cancellation: while the large-scale field can be recovered, small-scale components are lost, leading to a systematic underestimation of the available magnetic energy budget. To mitigate this limitation, some studies adopt an empirical correction by applying a uniform multiplicative factor to the ZDI-derived field strength (e.g., a five-fold enhancement in D. Evensberget et al. 2021, 2022, 2023). While this approach can partially compensate for the missing flux, it remains an approximation that does not fully capture the complexity of unresolved small-scale structures. Models by J. D. Alvarado-Gómez et al. (2016a) show that using ZDI maps to model stellar coronae can significantly underestimate the XUV radiative flux, with discrepancies reaching up to 1–2 orders of magnitude. As a result, most of the current 3D MHD models that utilize ZDI magnetograms are primarily focused on simulating large-scale stellar winds and the interplanetary environment, rather than modeling coronal structures.

[10] For comparison, the Sun exhibits an X-ray luminosity of approximately $10^{27}$ erg s$^{-1}$ in the 0.1–2.4 keV passband and an $L_X/L_{bol}$ ratio in the range of $10^{-7}$–$10^{-6}$ (P. G. Judge et al. 2003).

[11] The nearest main-sequence star is known to be Proxima Centauri, located approximately 4.2 lt-yr from the Sun.

Another branch of getting stellar magnetic fields is global convective dynamo (hereafter referred as GCD) simulations, in which the bulk of the convection zone is modeled using the full MHD equations (for a recent review, see P. J. Käpylä et al. 2023). This type of model investigates the dynamo processes in the convection zone, comprising the large-scale dynamo (LSD; see, e.g., P. Charbonneau 2014; P. Charbonneau & D. Sokoloff 2023) and the small-scale (or fluctuation) dynamo (SSD; for a recent review, see M. Rempel et al. 2023). These models aim to self-consistently generate and capture the large-scale flows, namely differential rotation and meridional circulation, and small-scale turbulence under the influence of rotation and stratification, important both for LSD and SSD. Compared with ZDI, GCD simulations have the advantage of resolving small-scale structures in addition to the large-scale field features. Recently, M. Viviani et al. (2018) reported on such a suitable set of GCDs to feed the stellar coronal simulations via boundary conditions as described later.

In this study, instead of ZDI maps, we employed surface magnetic maps for four solar-type stars with different rotation rates (1 $M_\odot$, 1 $R_\odot$, 1.8–23.3 $\Omega_\odot$[12]) from GDC models of M. Viviani et al. (2018). For each star, we selected two representative magnetic maps corresponding to magnetic activity minimum and maximum phases, respectively. To facilitate comparison with the Sun (1.0 $\Omega_\odot$, $P = 25.35$ days), we also utilized two solar magnetograms from the Helioseismic and Magnetic Imager (HMI; J. Schou et al. 2012) on board the Solar Dynamics Observatory (SDO; W. D. Pesnell et al. 2012). By accounting for small-scale magnetic flux, we construct a more realistic representation of the coronal structures and X-ray emissions of solar-type stars across different rotation periods and activity levels. Our research also serves as a foundation for future efforts to understand the dynamics of transients such as stellar flares and coronal mass ejections (CMEs). This paper is organized as follows: Section 2 introduces our numerical methodology, Section 3 reports the key model results of stellar coronae, Section 4 is a discussion, and Section 5 is a conclusion. This paper forms the first part of a two-part study. In Paper II, we will turn our attention to the properties and dynamics of stellar winds derived from the same set of models.

## 2. MHD Model and Input

### 2.1. Alfvén Wave Solar Atmosphere Model (AWSoM)

In this study, we utilized AWSoM (B. van der Holst et al. 2010; I. V. Sokolov et al. 2013; B. van der Holst et al. 2014; X. Meng et al. 2015; B. van der Holst et al. 2022) to simulate stellar coronae and stellar winds within the SWMF (G. Tóth et al. 2005, 2012; T. I. Gombosi et al. 2021). The validation of AWSoM has been performed by comparing with a variety of observational data sets, such as extreme ultraviolet (EUV) images that capture the structure of coronal holes and active regions (I. V. Sokolov et al. 2013; B. van der Holst et al. 2014), in situ solar wind measurements (R. Oran et al. 2013; X. Meng et al. 2015; L. K. Jian et al. 2016; N. Sachdeva et al. 2019, 2021, 2023; B. van der Holst et al. 2022), white-light images of CMEs (W. B. Manchester et al. 2012, 2014; M. Jin et al. 2017a, 2017b), and coronal loop structures (T. Shi et al. 2022, 2024).

[12] $M_\odot = 1.989 \times 10^{33}$ g, $R_\odot = 6.957 \times 10^{10}$ cm, $\Omega_\odot = 2.865 \times 10^{-6}$ rad s$^{-1}$.

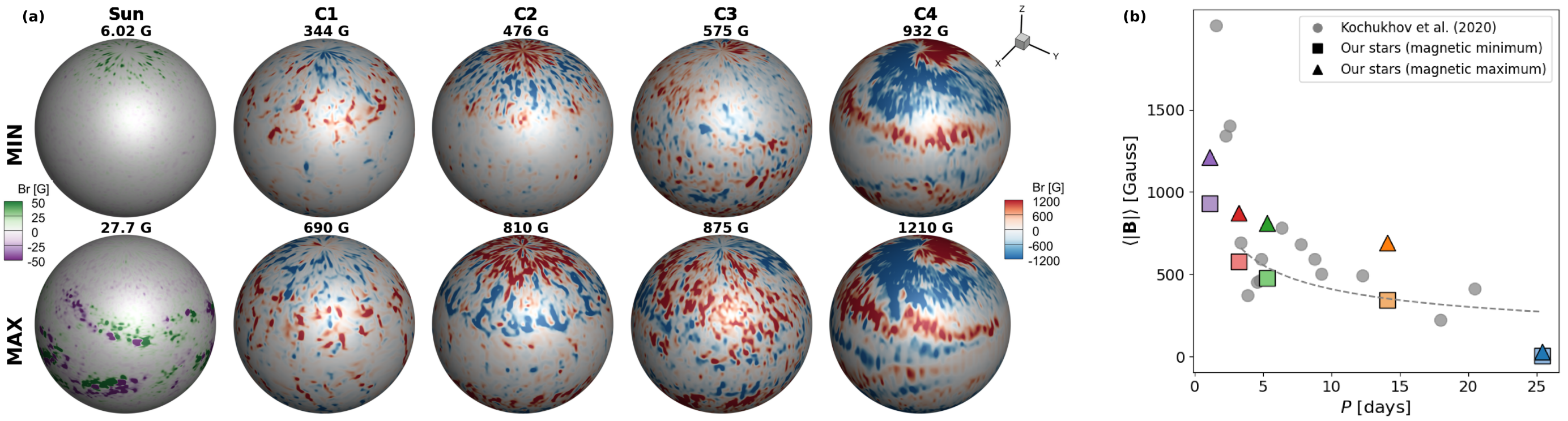

**Figure 1.** Magnetic boundary conditions in terms of topology (panel (a)) and field strength (panel (b)) in our models. (a) Surface radial magnetic field ($B_r$) in our models. The first column shows solar radial magnetograms from SDO/HMI, while the remaining columns display dynamo-generated radial maps from M. Viviani et al. (2018), scaled by a factor of 1/3 to match the magnetic field strengths observed in solar-type stars. The top and bottom rows represent the stellar activity minimum and maximum phases, respectively. Negative $B_r$ is shown in blue, while positive $B_r$ is shown in red. (b) Colored squares and triangles indicate mean magnetic field strengths of our stars at their minimum and maximum activity levels, respectively. The gray dots represent observed values by employing Zeeman broadening with masses between 0.7–1.3$M_\odot$ and radii between 0.7–1.3$R_\odot$ (O. Kochukhov et al. 2020). The gray dashed line shows our power-law fit for their nonsaturated stars. An animation with a duration of 23 s is available, which shows the surface distribution of the radial magnetic field for stars at different rotation rates.

(An animation of this figure is available in the online article.)

AWSoM is a 3D global extended MHD model, built on BATSRUS (K. G. Powell et al. 1999; T. I. Gombosi et al. 2004) that solves the full set of MHD equations in the heliographic rotating frame (HGR), beginning with the upper chromosphere and extending throughout the corona. It adopts the transformation method proposed by R. Lionello et al. (2009), which helps preserve steep gradients in temperature and density near the transition region. It uses a physics-based heating formulation instead of ad hoc heating functions. At the lower boundary (the solar/stellar surface), outward-propagating Alfvén waves are empirically injected via the free input parameter $S_A/B$, where $S_A$ is the Poynting flux and $B$ is the local magnetic field strength. Counter-propagating Alfvén waves nonlinearly interact, driving turbulent cascades that dissipate wave energy and thereby heat the plasma and accelerate the solar wind (M. Velli et al. 1989; G. P. Zank et al. 1996, 2012; W. H. Matthaeus et al. 1999; B. D. G. Chandran et al. 2011). In this work, we adopted a simplified two-temperature model assuming isotropic temperatures for both protons and electrons, and their respective heating rates are partitioned from the total turbulent heating according to the prescriptions of Y. Lithwick et al. (2007) and B. van der Holst et al. (2022). The electron heat flux $q_e$ transitions from the collision heat conduction formulation of L. Spitzer & R. Härm (1953) to the collisionless heat flux by J. V. Hollweg (1978) at around $r = 5R_\star$, as introduced by B. van der Holst et al. (2014). Optically thin radiative losses are computed using emissivities from the CHIANTI atomic database (K. P. Dere et al. 1997; R. P. Dufresne et al. 2024).

### 2.2. Input Magnetic Field

In the GCD simulations by M. Viviani et al. (2018), all four stars are simulated with the same mass ($M_\star = 1\,M_\odot$) and radius ($R_\star = 1\,R_\odot$) but differ in rotation periods, which makes them well suited for studying the effects of stellar rotation on stellar coronae and winds in solar-type stars. We selected four stars: C1 ($1.8\,\Omega_\odot$, $P = 14.1$ days), C2 ($4.8\,\Omega_\odot$, $P = 5.28$ days), C3 ($7.8\,\Omega_\odot$, $P = 3.25$ days), and C4 ($23.3\,\Omega_\odot$, $P = 1.09$ days). The simulations span decades to a century of evolution for each star, from which we selected two magnetic maps corresponding to the maximum and minimum $\langle|B_r|\rangle$ (mean unsigned radial magnetic field strength) to represent the minimum and maximum activity levels, respectively.

We preprocessed the magnetic maps in two steps. First, we extracted the magnetic field distribution from the simulations. The GCD simulations do not resolve the stratification or magnetic structure of the stellar photosphere, as radiative transfer is not included due to computational constraints. In addition, instead of modeling the stellar surface directly, the simulations impose a physically motivated boundary condition that enforces the magnetic field to be radial near the surface. As a result, we used the magnetic field at $r = 0.98\,R_\star$ rather than exactly at the surface ($r = R_\star$), since the boundary effects are significantly reduced at this depth, yielding a more physically meaningful magnetic field structure. Furthermore, the GCD simulations cover latitudes from $-75°$ to $+75°$, omitting the polar regions (see Figure 1 in M. Viviani et al 2018). To account for this, we extended the magnetic maps to the full latitude range of $-90°$ to $+90°$ using linear extrapolation, following the method described in T. Hackman et al. (2024). The resulting full-sphere map is shown in Figure 1(a). Global solar simulations using the AWSoM model have demonstrated that magnetograms with angular resolutions of approximately 1°–2° are sufficient to capture the essential physical features of the solar corona. The magnetograms derived from the GCD simulations resolve magnetic structures down to ∼2°, which is consistent with the resolution typically used in current global MHD modeling. Note that the C2 run covered only $\pi/2$ in longitude but was replicated to $2\pi$, which means only azimuthal order $m = 4$ and its multiples are present in the spherical harmonic expansion. The periodic repetition of C2 in longitude is precisely due to this reason.

After the above processing, the field structures for solar-type stars are determined, but the strength is overestimated relative to the expected values. Here, we rescaled the magnetic field strength. Specifically, we adopted empirical relations between stellar rotation rates and surface-averaged field strength $\langle|\boldsymbol{B}|\rangle$ for solar-type stars to set the scaling factor. In Figure 1(b), we plotted the G-type stars from the work by O. Kochukhov et al. (2020), whose $\langle|\boldsymbol{B}|\rangle$ are measured by Zeeman broadening

**Table 1**
Key Parameters of Inputs and Outputs of Our Modeled Stars

| Case | $Sun_{min}$ | $Sun_{max}$ | $C1_{min}$ | $C1_{max}$ | $C2_{min}$ | $C2_{max}$ | $C3_{min}$ | $C3_{max}$ | $C4_{min}$ | $C4_{max}$ |
|---|---|---|---|---|---|---|---|---|---|---|
| $P$ [d] | 25.38 | 25.38 | 14.10 | 14.10 | 5.29 | 5.29 | 3.25 | 3.25 | 1.09 | 1.09 |
| $\Omega[\Omega_\odot]$ | 1.0 | 1.0 | 1.8 | 1.8 | 4.8 | 4.8 | 7.8 | 7.8 | 23.3 | 23.3 |
| $f_n$ | 1 | 1 | 2.84 | 7.41 | 4.42 | 9.32 | 5.75 | 10.26 | 11.86 | 16.86 |
| $S_A/B$ [MW m$^{-2}$ T$^{-1}$] | 1 | 1 | 1.69 | 2.72 | 2.10 | 3.05 | 2.40 | 3.20 | 3.44 | 4.11 |
| $\langle\lvert\boldsymbol{B}_{\rm initial}\rvert\rangle$ [G] | 4.93 | 24.8 | 316 | 642 | 438 | 761 | 532 | 817 | 909 | 1180 |
| $\langle\lvert\boldsymbol{B}\rvert\rangle$ [G] | 6.02 | 27.7 | 344 | 690 | 476 | 810 | 575 | 875 | 932 | 1210 |
| $n_{1.5R_\star}$ [cm$^{-3}$] | 5.41E+06 | 2.52E+07 | 2.45E+08 | 4.99E+08 | 2.04E+08 | 5.04E+08 | 3.47E+08 | 5.14E+08 | 6.07E+08 | 7.61E+08 |
| $L_{\rm X}$ [erg s$^{-1}$] | 3.23E+26 | 2.19E+27 | 1.03E+29 | 3.92E+29 | 1.01E+29 | 3.64E+29 | 2.63E+29 | 5.30E+29 | 5.11E+29 | 7.31E+29 |
| $F_{\rm X}$ [erg s$^{-1}$ cm$^{-2}$] | 5.31E+03 | 3.60E+04 | 1.69E+06 | 6.44E+06 | 1.66E+06 | 5.98E+06 | 4.32E+06 | 8.71E+06 | 8.40E+06 | 1.20E+07 |

**Note.** $f_n = \frac{n_\star}{n_\odot}$ represents the initial density ratio of the stellar atmosphere relative to that of the Sun. $\langle|\boldsymbol{B}_{\rm initial}|\rangle$ denotes the mean magnetic field strength as the initial input, evaluated at $r = 1.001R_\star$, while $\langle|\boldsymbol{B}|\rangle$ represents the mean magnetic field strength after relaxation, when the magnetic field is fully coupled with the plasma. $n_{1.5R_\star}$ denotes the average electron number density on the spherical surface at $r = 1.5R_\star$.

methods. We did a fit to their data as plotted by the gray dashed line. To ensure stars in the unsaturated regime (C1, C2, C3) align with the gray line, we applied a scaling factor of 1/3 to the raw GCD data and finally got the $\langle|\boldsymbol{B}|\rangle$ for our solar-type stars as shown in Figure 1(b). Unlike C1–3, the rotation period associated with C4 would make it fall within the saturated regime. The solar maximum case (CR2287) and solar minimum case (CR2223) from SDO/HMI are located below the trend line.

In AWSoM, we input the radial magnetic field $B_r$, and then used the potential field source surface (PFSS) extrapolation to derive the azimuthal ($B_\phi$) and poloidal ($B_\theta$) components, from which we computed surface mean magnetic field strength $\langle|\boldsymbol{B}|\rangle$ at $r = 1.001R_\star$ for each star. As the magnetic fields fully couple with plasma, they change from $\langle|\boldsymbol{B}_{\rm initial}|\rangle$ to final $\langle|\boldsymbol{B}|\rangle$, as listed in Table 1. These changes are negligible and do not affect the overall results or conclusions of our model.

## 2.3. Setup

We used spherical grids within the solar corona (SC) component of the SWMF. The inner boundary was set at 1 $R_\star$ (1 solar radius). To ensure that the Alfvén surfaces (AS) were located inside the model domain, the outer boundary was set at 80 $R_\star$ for C4 and 40 $R_\star$ for the other stars. The domain was decomposed into grid blocks with $6 \times 8 \times 8$ internal grid cells in the radial, longitudinal, and latitudinal directions, respectively. At the base level, we employed $24 \times 16 \times 8$ adaptive mesh refinement (AMR) blocks for C4 and $16 \times 16 \times 8$ AMR blocks for the other stars. The base resolution was $2\overset{\circ}{.}8$ in both the longitudinal and latitudinal directions. The radial grid spacing, $\Delta r$, was defined as a smooth function of $\ln(r)$ and became uniform in the outer corona. To better resolve the lower corona at $r < 2R_\star$, we applied two additional levels of AMR for the solar-type stars. For the Sun, we adopted a three-level AMR configuration following the work by T. Shi et al. (2022), which enabled the model to more accurately capture hot, dense active regions and reproduce X-ray emission consistent with observations (see Section 3.3). With this setup, the smallest radial grid spacing reached $\Delta r \approx 0.0003$–$0.0004$ $R_\odot$ (approximately 210–280 km), and the angular resolution at $r < 2R_\star$ improved to $0\overset{\circ}{.}35$ for the Sun and $0\overset{\circ}{.}7$ for solar-type stars. The total number of cells in the SC component was 121 million for the Sun, 18 million for C1–3, and 23 million for C4.

For active stars with stronger magnetic fields (400–1300 G), we expect their coronae to be denser. To allow the corona to reach the expected density, the base densities are scaled proportionally. To do this, we first estimated the X-ray luminosity ($L_{\rm X}$) from $\langle|\boldsymbol{B}|\rangle$ using the scaling laws for G-type stars from the work by O. Kochukhov et al. (2020), i.e., $\log(L_{\rm X}/L_{\rm bol}) = 2.7\log\langle|\boldsymbol{B}|\rangle - 12.1$ with $L_{\rm bol} = 3.828 \times 10^{33}$ erg s$^{-1}$. Assuming $L_{\rm X} \propto n^2$, the stellar-to-solar density scaling factor is then set as $f_n = n_\star/n_\odot \propto (L_{{\rm X},\star}/L_{{\rm X},\odot})^{0.5}$, where $L_{{\rm X},\odot} = 2.24 \times 10^{27}$ erg/s (N. J. Wright et al. 2011). The values of $f_n$ are listed in Table 1. The initial density for the solar cases, $n_\odot$, follows the setting by N. Sachdeva et al. (2019). As the MHD equations evolve toward a relaxed steady state, the atmospheric temperature and density naturally adjust to a physically consistent structure.

The wave energy density of the input Alfvén wave is determined by the Poynting flux ($S_{\rm A}$) of the outward-propagating wave at the inner boundary. In AWSoM, $S_{\rm A}$ is set to be proportional to the magnetic field strength $B$, and the ratio $S_{\rm A}/B$ is treated as an adjustable parameter (I. V. Sokolov et al. 2013; B. van der Holst et al. 2014). This setup allows higher wave energy to be injected to regions with stronger magnetic fields, which is consistent with the physical expectation of Alfvén-wave-induced coronal heating. In this paper, we use $S_{\rm A}/B = 1.0$ MW m$^{-2}$ T$^{-1}$ for the Sun (I. V. Sokolov et al. 2013; N. Sachdeva et al. 2019, 2021, 2023; Z. Huang et al. 2023). However, for the other stars, to heat the coronae to a temperature close to the observations (see Section 3.2), it is necessary to adjust $S_{\rm A}/B$ to maintain a reasonable coronal heating rate. The choice of how to prescribe $S_{\rm A}$ is not unique (e.g., S. R. Cranmer & S. H. Saar 2011). Here, we use a simplified model to help us adjust the value of $S_{\rm A}/B$. The Poynting flux, $S_{\rm A}$, is given by $S_{\rm A} = \varepsilon_{\rm wave}V_{\rm A}$, where $\varepsilon_{\rm wave}$ is the wave energy density and $V_{\rm A} = B/\sqrt{\mu_0\rho}$ is the Alfvén velocity. At the inner boundary, the wave energy density $\varepsilon_{\rm wave}$ can be expressed in terms of plasma density $\rho$ and velocity fluctuations $v'^2$ as $\varepsilon_{\rm wave} \propto \rho v'^2$. Assuming similar velocity fluctuations $v'$ in solar-type stars (S. Boro Saikia et al. 2023), the $S_{\rm A}/B$ scales approximately as $\rho^{0.5}$. Since plasma density $\rho$ is set based on assumed X-ray luminosity $L_{\rm X}$, the scaling of $S_{\rm A}/B$ is inherently constrained by assumed $L_{\rm X}$. A more sophisticated prescription of $S_{\rm A}/B$ would require further investigation of coronal heating and observations, which is beyond the scope of this work. The values of $S_{\rm A}/B$ in this work are listed in Table 1.

Local time stepping (G. Tóth et al. 2012) is used for speeding up the convergence of the simulation to a steady state. We consider the simulation converged when, over 10,000 iterations, the X-ray luminosity ($L_X$) changes by less than 5%, and the mean coronal density and temperature over spherical shells at 1.5 $R_\star$, 5 $R_\star$, and 10 $R_\star$ change by less than 1%. Each case typically requires a total of 120,000–160,000 iterations to reach convergence.

## 3. Results

### 3.1. Coronal Structure

Figure 2 presents the distribution of the electron number density ($n$) via color maps on the $xz$-plane and the electron temperature ($T_e$) using contour maps for each star. The apparent symmetry in the case of C2 arises from the construction of the input magnetogram, which was generated by replicating a $\pi/2$ longitude segment four times (see Section 2.2).

The electron number density $n$ exhibits a monotonic decline with increasing radial distance. At $r \sim 1.5\,R_\star$, corresponding to a height of $0.5\,R_\star$ above the stellar surface, we calculate the average electron number density over the spherical surface, denoted as $n_{1.5\star}$, as listed in Table 1. The typical value of $n_{1.5\star}$ for active solar-type stars is approximately $2\text{–}8 \times 10^8\,\text{cm}^{-3}$, which is 1–2 orders of magnitude higher than the solar coronal density of $\sim 1 \times 10^7\,\text{cm}^{-3}$. The $n_{1.5\star}$ represents an average over both high- and low-density regions. As shown in the coronal structures in Figure 2, localized high-density active regions at $r \sim 1.5\,R_\star$ can exhibit electron number densities exceeding $1 \times 10^9\,\text{cm}^{-3}$. These regions are also characterized by ultra-hot temperatures.

In the solar cases, $T_e$ typically rises from transition region values of approximately $10^5$ K to peak coronal temperatures of $1\text{–}4 \times 10^6$ K. For stars with stronger magnetic fields of 300–1200 G, the peak coronal temperatures can reach values between $1 \times 10^7$ K and $2 \times 10^7$ K, as indicated by the red and orange contours. These high-temperature regions also extend spatially. For instance, the temperature of $7.2 \times 10^6$ K, as represented by the orange contours, extends to the radii of 3 $R_\star$ (i.e., in the case of C1$_{\text{max}}$). Furthermore, in magnetically active cases such as C4$_{\text{max}}$, the temperature of $1 \times 10^7$ K even appears in a more extended region, reaching radii exceeding 7 $R_\star$ in the polar region, suggesting the hottest coronal region is much higher than that in the solar cases.

Both the high-density and high-temperature features are consistently associated with closed magnetic field lines, as illustrated in the second and fourth columns of Figure 2 and the X-ray active region as discussed in Section 3.3.3. In contrast, coronal regions that exhibit relatively low densities and temperatures are associated with open magnetic fields and manifest as coronal holes in X-ray imaging.

### 3.2. Emission Measure Distribution

To validate our coronal models, we conducted an EMD analysis following the volume emission measure definition of N. S. Brickhouse & A. K. Dupree (1998), where

$$\text{EM}_V(T_i) = \int_{V(T-\frac{\Delta T_i}{2} \leqslant T < T+\frac{\Delta T_i}{2})} n_e n_\text{H}\, dV$$
$$= \int_{V(T-\frac{\Delta T_i}{2} \leqslant T < T+\frac{\Delta T_i}{2})} n^2\, dV,$$

under the assumption of quasi-neutrality ($n_e \approx n_\text{H}$). $T - \frac{\Delta T_i}{2}$ and $T + \frac{\Delta T_i}{2}$ denote the lower and upper bounds of the $i$th temperature bin. The bins are defined uniformly in logarithmic temperature space, with a width of $\Delta \log T = 0.1$. Our results are presented as colored lines in Figure 3.

We compared our results with predictions from B. E. Wood (2018), which conducted a statistical analysis of a large sample of stars and provided typical EMDs for different levels of surface X-ray flux ($F_X$, in erg cm$^{-2}$ s$^{-1}$, in the 0.1–2.4 keV band). In our figure, we include these EMDs for $\log F_X = 4.0$, 5.0, 6.0, and 7.0 as reference curves, with shaded regions between them (from dark to light gray) indicating EMD in increasing $F_X$ levels.

Figure 3 also includes observational constraints from four stars. Two are active M dwarfs: AD Leo (M3.0V, $R_o \approx 0.047$, $\log F_X \approx 6.8$, $\langle|\boldsymbol{B}|\rangle \approx 3.6$ kG) and AU Mic (M0.5V, $R_o \approx 0.14$, $\log F_X \approx 7.1$, $\langle|\boldsymbol{B}|\rangle \approx 2.6$ kG), both their EMD are from J. Sanz-Forcada et al. (2025). The other two are solar analogs: ι Horologii (F8V/G0V, $R_o \approx 0.89$, $\log F_X \approx 5.85$, $\langle|\boldsymbol{B}|\rangle \approx 0.5$ kG) whose EMD is from J. Sanz-Forcada et al. (2019), and HIP 67522 (G0V, $R_o \approx 0.15$, $\log F_X \approx 7.42$, $\langle|\boldsymbol{B}|\rangle \approx 1.6$ kG) whose EMD is from A. Maggio et al. (2024).[13]

The difference between our Sun$_{\text{max}}$ and Sun$_{\text{min}}$ models arises from the number of active regions, where the highest temperature could exceed $10^{6.5}$ K. Compared to the solar cases, there are ultra-hot plasmas with temperatures $T_e > 10^{6.8}$ K in solar-type stars, and the overall EM$_V$ values also increase. For these stars, we analyze three temperature regimes: the ultra-hot regime ($T_e > 10^{6.8}$ K), the warm regime ($T_e < 10^{6.2}$ K), and the hot regime in between.

In the ultra-hot regime ($T_e > 10^{6.8}$ K), our simulated solar-type stars exhibit high emission measure ($\text{EM}_V = 10^{50}$–$10^{52}\,\text{cm}^{-3}$). Most of the X-ray emission originates from these ultra-hot plasmas. Observations suggest that stars with higher X-ray surface flux ($F_X$) tend to exhibit stronger EMs in the ultra-hot regime. This trend is reflected in the shaded regions of our figure, which are based on the EMDs from B. E. Wood (2018): the increasingly lighter gray shades represent higher $\log F_X$ levels, with EMDs systematically shifting upward with increasing $F_X$. In our models, most solar-type stars have $6.0 < \log F_X < 7.0$, with only C4$_{\text{max}}$ reaching $\log F_X = 7.1$ (see Section 3.3 and Table 1). When compared with the shaded region corresponding to $\log F_X = 6.0\text{–}7.0$, our simulated EMDs show good agreement with the observational trend, supporting the physical plausibility of the high-temperature coronal structures captured in our models. Besides, we also include two individual solar analogs for reference in Figure 3. The lower and upper dashed lines represent ι Horologii ($\log F_X \approx 5.85$) and HIP 67522 ($\log F_X \approx 7.42$), respectively. Our results lie between these two extremes, further illustrating the consistency of the modeled EMDs with observed stellar coronae across a range of activity levels.

In the hot regime ($10^{6.2}$ K $< T_e < 10^{6.8}$ K), our results show similar slopes to the EMDs of AD Leo and AU Mic, as reconstructed by J. Sanz-Forcada et al. (2025). Both the stars show $\log F_X$ of around 7.0, which are comparable to the values of our sample stars, suggesting a physical similarity despite their significantly stronger $\langle|\boldsymbol{B}|\rangle$ (3–4 kG). Again, the results of our model are bounded by the curves of solar analogs ι Horologii and HIP 67522, which further supports the model's applicability to solar-type stars.

[13] The values of $R_o$ and $\langle|\boldsymbol{B}|\rangle$ for AD Leo and AU Mic are taken from J. Morin et al. (2008) and J. F. Donati et al. (2025), respectively. For ι Horologii and HIP 67522, the $R_o$ are estimated using Equation (6) from N. J. Wright et al. (2018), and the corresponding $\langle|\mathbf{B}|\rangle$ values are calculated based on the $R_o$–$\langle|\boldsymbol{B}|\rangle$ relation given in Equation (3) of O. Kochukhov et al. (2020).

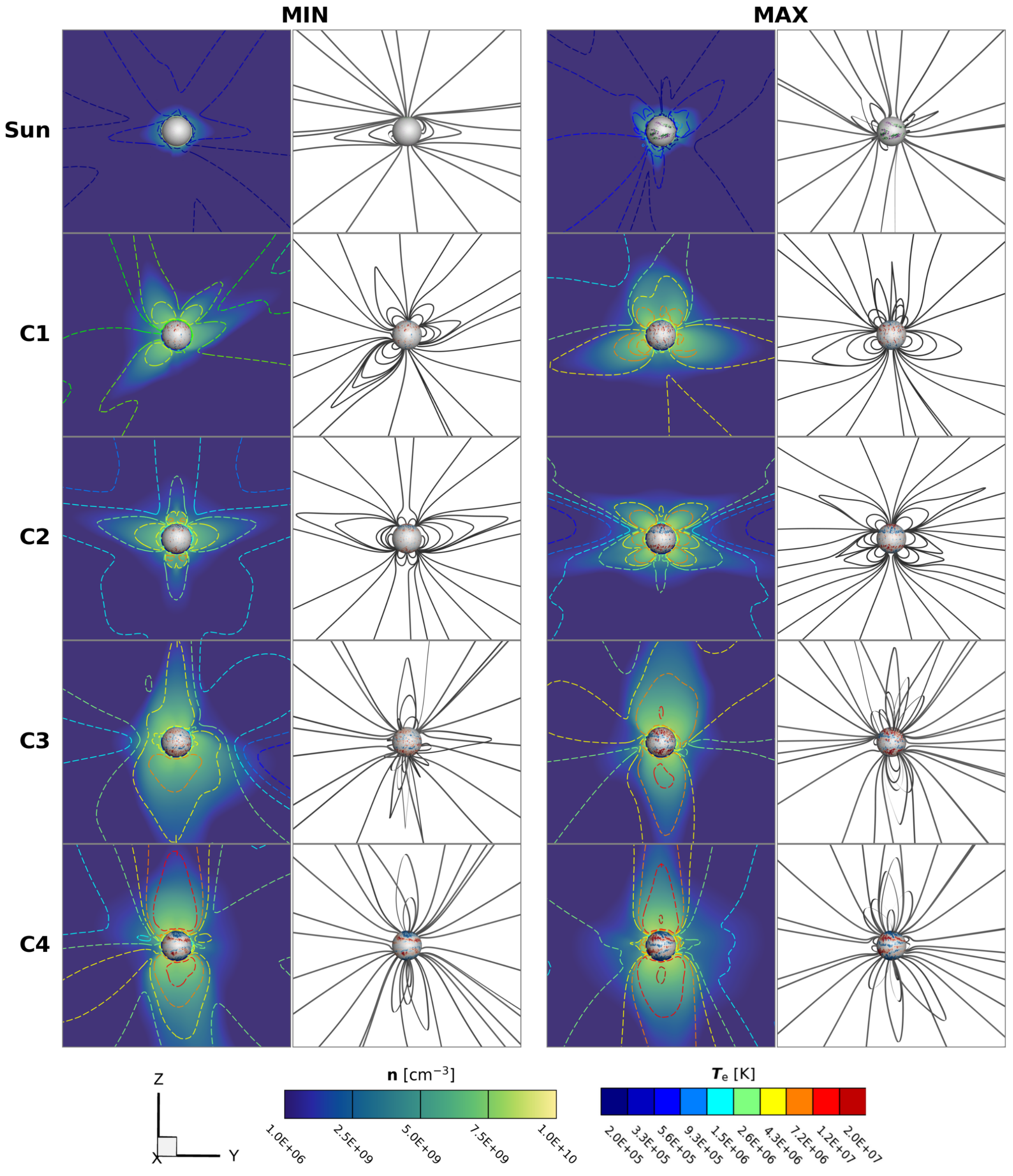

**Figure 2.** Simulated stellar coronal structures at the *yz*-plane. The first and second columns correspond to the activity minimum phase, while the third and fourth columns represent the activity maximum phase. In each case, the first and third columns show the distribution of electron temperature ($T_e$) as contour maps and electron number density ($n$) as color maps, both plotted on the *yz*-plane for each star. The second and fourth columns display the corresponding magnetic field lines, which are traced from seed points located only in the *yz*-plane. The magnetograms used are the same as those presented in Figure 1.

In the warm regime ($T_e < 10^{6.2}$ K), our models overpredict $EM_V$ by more than 1 dex. A similar excess is also presented by Y. Xu et al. (2025), where they simulated a young solar-type star (ι Horologii) using AWSoM. Most of this excess originates from the lower corona, which is likely due to the code's difficulty in accurately capturing the steep plasma density gradient between the transition region and the corona. The model tends to overestimate the plasma density there, particularly at temperatures of $T < 10^{6.2}$ K, resulting in enhanced EM in the range. However, as the height increases, such an influence will diminish.

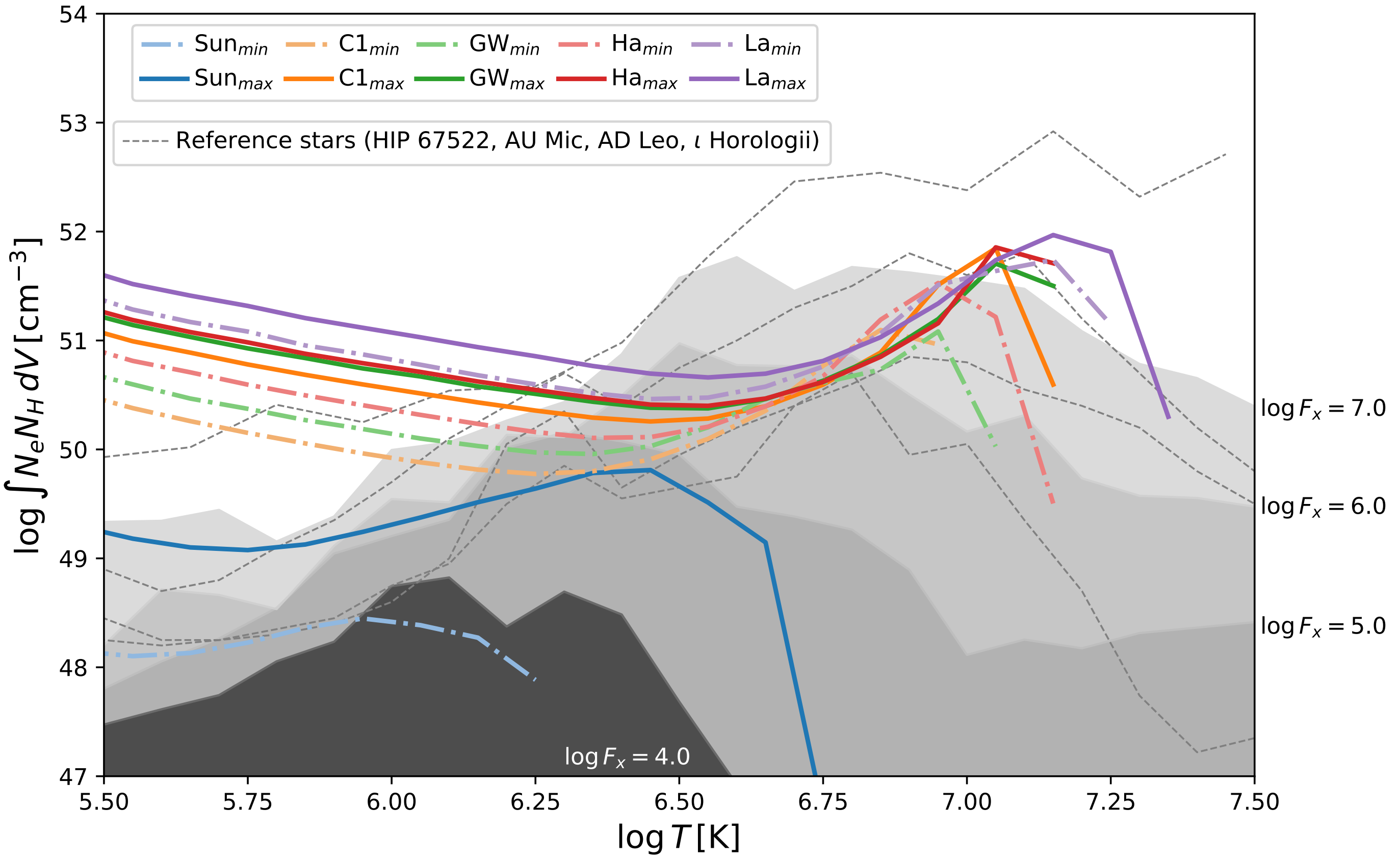

**Figure 3.** Comparison of observed and simulated emission measure distributions (EMDs). Colored lines represent our model results. For each color, the dark solid lines and light dashed–dotted lines indicate cases at maximum and minimum activity levels, respectively. The dashed gray lines show the reconstructed EMDs based on observations of HIP 67522 (A. Maggio et al. 2024), AU Mic (J. Sanz-Forcada et al. 2025), AD Leo (J. Sanz-Forcada et al. 2025), and ι Horologii (J. Sanz-Forcada et al. 2019), plotted from top to bottom according to their overall emission measure values. The gray shaded regions indicate predicted EMDs for stars with $\log F_X$= 4.0–7.0 based on B. E. Wood (2018). The shading from dark to light gray corresponds to increasing $F_X$.

### 3.3. Soft X-Ray Emissions

#### 3.3.1. Soft X-Ray Spectrum

Continuum intensities and spectral lines were calculated using the CHIANTI Spectral Synthesis Package (v11.0; K. P. Dere et al. 1997; R. P. Dufresne et al. 2024). A detailed description of the calculation setup is provided in Appendix A. For solar-type stars C1–4, the $EM_V$ at $\log T < 6.2$ (Figure 3) is overestimated by 2–3 dex compared to observational constraints (e.g., B. E. Wood 2018), which would therefore lead to a significantly overestimated soft X-ray flux, especially in the energy band of below 0.3 keV. Their impact on the synthetic spectra is more substantial than the omission of their actual (but much smaller) contribution. To get a better approximation to the expected true spectra, we adopted the EMD excluding $\log T < 6.2$ plasma to compute synthetic spectra for solar-type stars C1–4. For the solar cases, we use the original EMD, given that it is well constrained. The results are shown in Figure 4.

In general, magnetically active stars with hotter coronal components exhibit higher luminosity density at high energies, and their high-energy spectra (above 2 keV) decline more slowly, with a harder spectral slope. The emission near 1.9 keV is enhanced by silicon lines, like Si XIII and Si XIV, while the spectral features around 3.9 keV are primarily contributed by Ca XVIII and Ca XIX. At approximately 6.6 keV, multiple iron lines appear, such as Fe XX– XXIV, and around 7.8 keV, Fe XXIV and Fe XXV become prominent. In the Sun, these lines are invisible except during solar flares (e.g., K. Tanaka et al. 1984) because they are typically formed at temperatures of $\log T$= 7.0–7.1. We suggest that these high-temperature emission lines can serve to diagnose the ultra-hot coronal plasma, as also demonstrated by previous observations (e.g., A. Maggio et al. 2024; M. Kurihara et al. 2025).

#### 3.3.2. Soft X-Ray Luminosity

Using the above spectra, we calculated the ROSAT PSPC 0.1–2.4 keV band luminosities (Figure 5). The colored points represent our results. Using the convective turnover time $\tau_c$ from the mass-dependent prescription of N. J. Wright et al. (2018;$\log \tau = 1.16 - 1.49 \log(M/M_\odot) - 0.54 \log^2(M/M_\odot)$), we computed the Rossby number $R_o = P_{rot}/\tau_c$. Assuming a fixed $L_{bol} = 3.828 \times 10^{33}$ erg s$^{-1}$ for all solar-type stars (A. Prša et al. 2016), we derived the X-ray to bolometric luminosity ratio $L_X/L_{bol}$ for our simulated cases. In Figure 5(a), the gray points correspond to observations of partially convective stars from N. J. Wright et al. (2011), with their best-fit trend shown as a dashed line ($L_X/L_{bol} = 8.68 \times 10^{-6}\,\mathrm{Ro}^{-2.18}$ for $\mathrm{Ro} > 0.13$, and $L_X/L_{bol} = 10^{-3.13}$ for $\mathrm{Ro} < 0.13$). Our results are in good agreement with the observational statistics reported by N. J. Wright et al. (2011). The C4 cases (purple points) reside in the X-ray saturated regime. Their $L_X/L_{bol}$ values are a bit lower than the black dashed line, because the magnetic fields of C4 are weaker than the typical magnetic field strength of stars with the corresponding $\mathrm{R_o}$, as shown in Figure 1(b). In contrast, the elevated $L_X/L_{bol}$ ratio for C1 (orange points) is attributed to its systematically larger $\langle|\boldsymbol{B}|\rangle$. Individual stars exhibit significant $L_x$ variations between their activity maximum and minimum. The variability contributes to the intrinsic scatter of $L_x$ at a given rotation period, beyond what is caused by differences between individual stars.

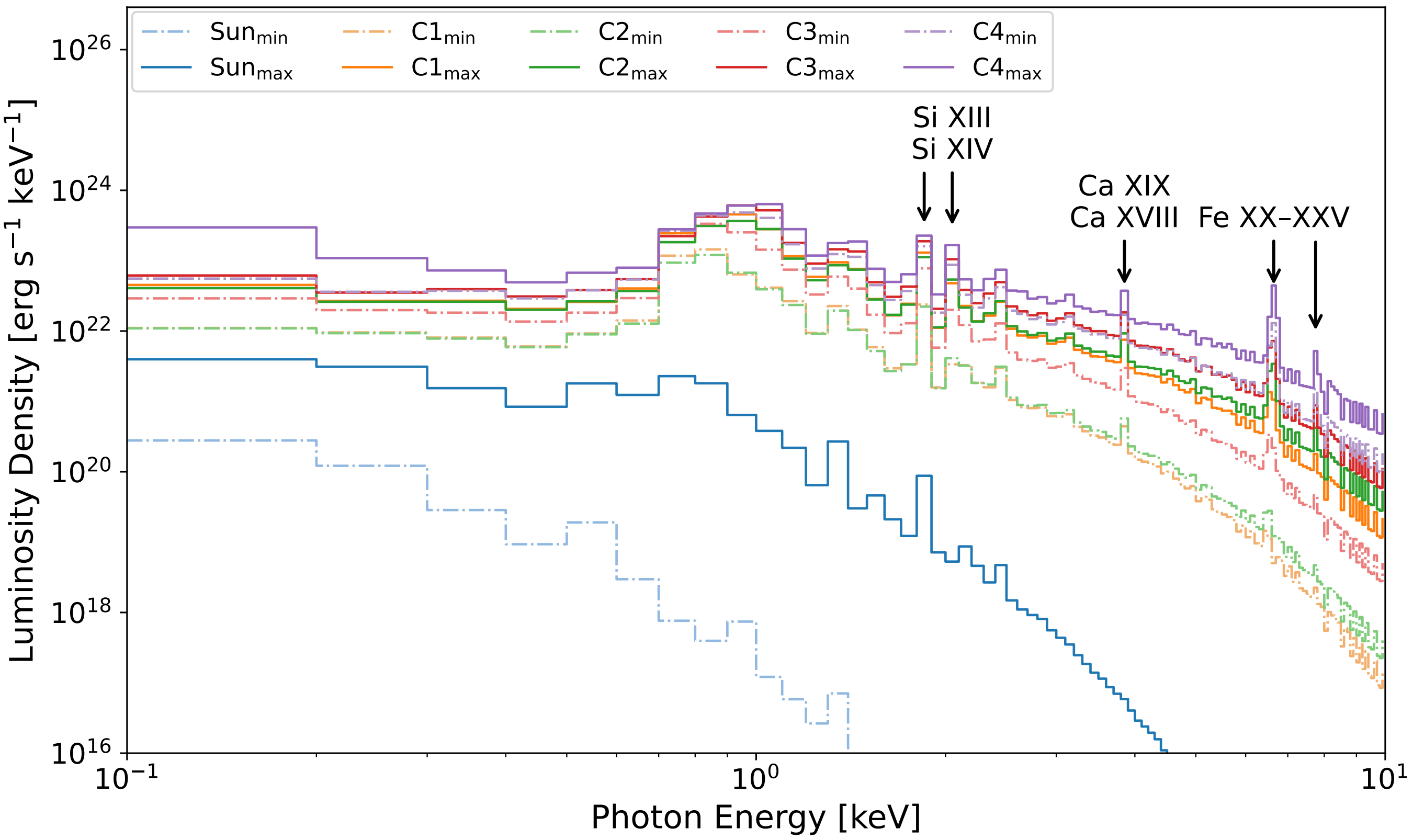

**Figure 4.** Synthetic X-ray spectra for our modeled stars and the Sun. The energy bin size is $\Delta E = 0.1$ keV. Several significant emission lines are also pointed out.

Figure 5(b) shows our results, and a log–log linear fit for our solar-type stars is shown by the red dashed line,

$$\log(L_X/L_{bol}) = (-9.07 \pm 1.40) + (1.75 \pm 0.23)\log\langle|\boldsymbol{B}|\rangle,$$

with a coefficient of determination ($R^2$) of 0.91. Fitting including the solar data yields $\log(L_X/L_{bol}) = (-8.31 \pm 0.05) + (1.48 \pm 0.05)\log\langle|\boldsymbol{B}|\rangle$. The gray dots in Figure 5(b) represent observational data from O. Kochukhov et al. (2020), while the gray line is their fitting result[14], $\log(L_X/L_{bol}) = (-12.08 \pm 0.70) + (2.70 \pm 0.15)\log\langle|\boldsymbol{B}|\rangle$. Our model results are broadly consistent with their statistical trend, although our fitted slope (1.75) is lower than theirs (2.70), which will be further discussed in Section 4.1.

[14] The original relation in O. Kochukhov et al. (2020) is given as: $\log\langle|\boldsymbol{B}|\rangle = (4.47 \pm 0.09) + (0.37 \pm 0.02) \cdot \log\left(\frac{L_X}{L_{bol}}\right)$. Here we invert the relation to express $\log(L_X/L_{bol})$ as a function of $\log\langle|\boldsymbol{B}|\rangle$, using standard error propagation to recalculate the uncertainties accordingly.

#### 3.3.3. *Soft X-Ray Imaging*

Figure 6 shows the X-ray imaging in the same logarithmic scale using the temperature response function of the thick beryllium (thick-Be) filter from Hinode/XRT (L. Golub et al. 2007). As detailed in Appendix B, the thick-Be filter provides a strong response in the high-temperature regime. The images are restricted to within 3 $R_\star$, encompassing the region where most of the X-ray emission is generated. Comparing Figure 6 and Figure 2, X-ray emissions are closely associated with closed magnetic field structures, making X-ray emission a reliable proxy for identifying closed field regions or active regions filled with dense and hot plasma. In contrast, the regions with predominantly open field lines or the coronal holes exhibit significantly lower X-ray emission. Note that synthesizing the individual emission strands as often seen in solar EUV and X-ray high-resolution images is still difficult, despite the spatial resolution already being rather high.

As the star rotates, the nonuniform distribution of these bright active regions and dark coronal holes exhibits a rotational modulation (S. Chandra et al. 2010; J. Sanz-Forcada et al. 2019). To investigate this modulation, we generated synthetic X-ray images for each of three inclination angles (90°, 60°, and 30°), where the inclination refers to the angle between the object's rotation axis and the observer's line of sight. For each inclination, we sampled 50 images uniformly over a full 360° rotational phase. The average intensity of each image was calculated to construct the synthetic X-ray light curves shown in the second and third rows of Figure 6. A stronger magnetic field leads to a stronger modulation because

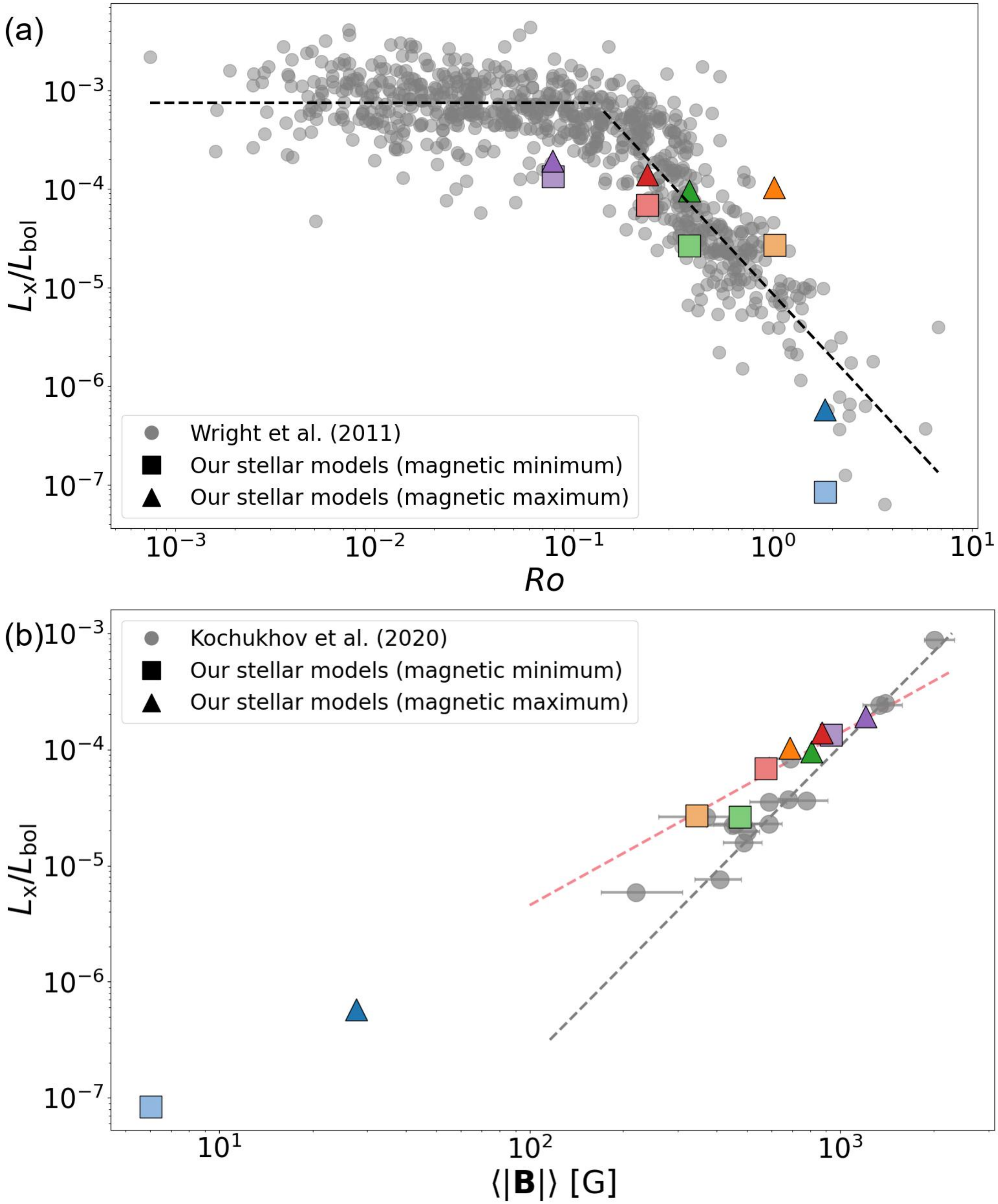

**Figure 5.** Stellar X-ray luminosity normalized by bolometric luminosity ($L_X/L_{bol}$) as a function of Rossby number (panel (a)) and mean surface magnetic field strength (panel (b)). Colored square and triangle markers represent model results, with colors consistent with Figure 4. Squares indicate the minimum cases, while triangles correspond to the maximum cases. A constant bolometric luminosity of $L_{bol} = 3.828 \times 10^{33}$ erg s$^{-1}$ is assumed for all simulated cases. Gray circles in panels (a) and (b) are observations taken from N. J. Wright et al. (2011) and O. Kochukhov et al. (2020), respectively, with the black and gray lines representing their empirical relations. The red dashed line shows a log–log linear fit to our solar-type data.

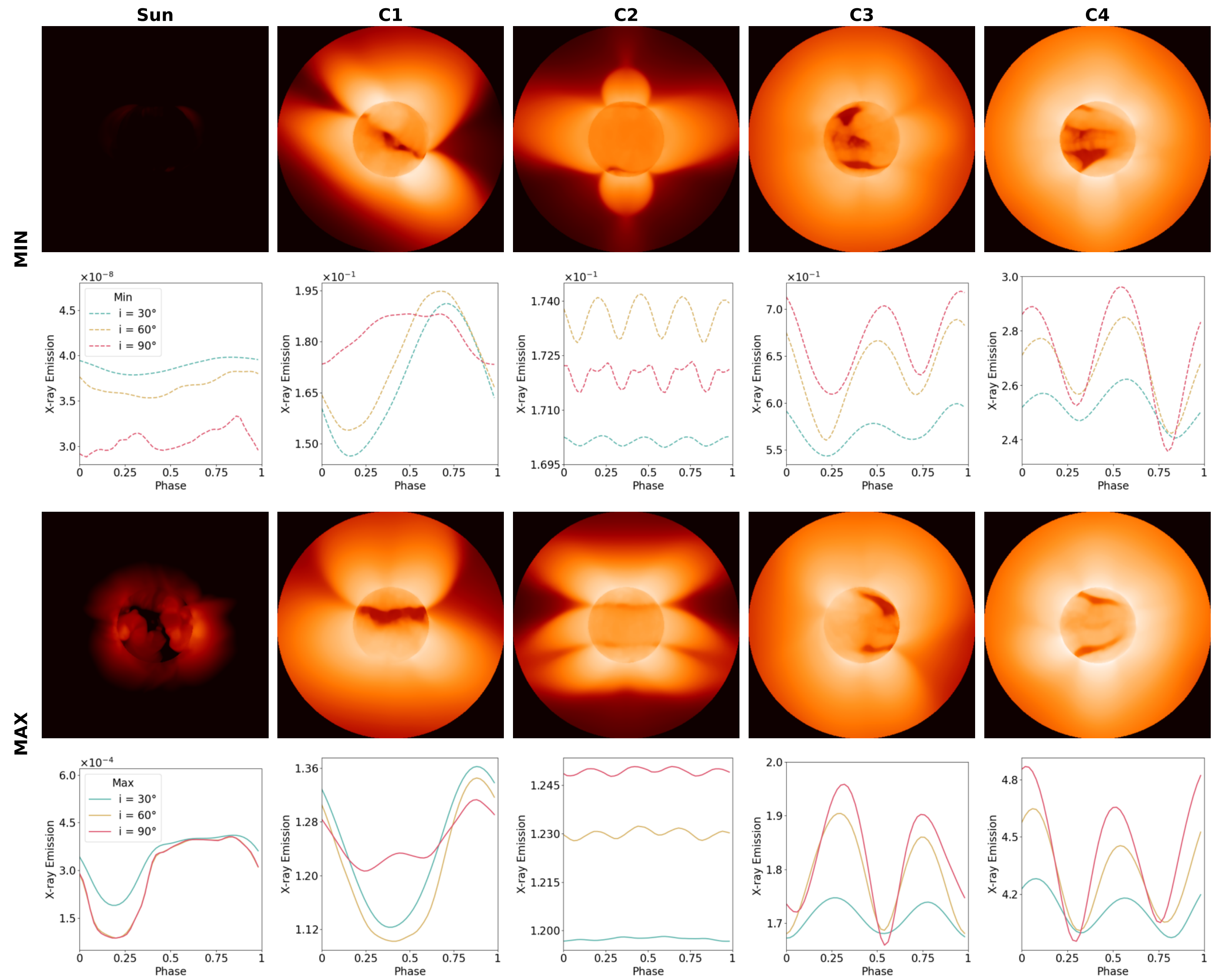

**Figure 6.** X-ray imaging and rotational modulation of the X-ray intensity. The first and third rows display synthetic Hinode/XRT X-ray images generated from an inclination angle of 90°, where the inclination is defined as the angle between the stellar rotation axis and the observer's line of sight. The second and fourth rows show the corresponding mean X-ray intensities as a function of rotational phase, computed for inclination angles of 30° (blue lines), 60° (yellow lines), and 90° (red lines). An animation with a duration of 3 s is available, which illustrates the X-ray emission from the rotating stars.

(An animation of this figure is available in the online article.)

the X-ray emission itself becomes more intense, resulting in a larger absolute variation. Our results suggest that, for solar-type stars, the X-ray rotational modulation can exceed 20% in certain cases, but the exact values depend on the viewing angle and magnetic configuration. Although viewing from the 90° inclination allows a larger area of the stellar surface to be visible, it does not necessarily correspond to the largest modulation amplitude. For instance, in the case of the $C1_{min}$ model, the rotational variation is more pronounced when viewed from higher latitudes.

## 4. Discussions

### 4.1. Scaling Relations Between Magnetic Field and X-Ray Luminosity

As shown in Figure 5(b), the $L_X/L_{bol}$–$\langle|\boldsymbol{B}|\rangle$ relation in our model is broadly consistent with the statistical trend from observations, although the fitted slope (1.75) is lower than 2.7, derived from observations by O. Kochukhov et al. (2020). Nonetheless, the slope we derived is in good agreement with several previously reported values that are in the range of 1.1–1.8 (e.g., A. A. Pevtsov et al. 2003; A. A. Vidotto et al. 2014). Given that magnetic energy scales as $B^2$, our power-law index of 1.75, smaller than 2, suggests that the increase in magnetic energy does not lead to a proportional increase in X-ray luminosity, indicating a limited efficiency in converting magnetic energy into X-ray radiation.

This result can be understood from how the Poynting flux $S_A$ is prescribed in our model. Specifically, for solar-type stars, we artificially set the input flux using the relation $\frac{S_A}{B} \propto n_{\rm initial}^{0.5} \propto B^{0.675}$, as discussed in Section 2.3, which reflects how the wave energy input scales with the base density and thus the magnetic field. Therefore, in our models, the Poynting flux scales as $S_A \propto B^{1.675}$. Assuming a linear conversion from the Poynting flux to radiative loss, it can be written: $L_X = \eta_{\rm rad} \cdot S_A \cdot A$, where $\eta_{\rm rad}$ denotes the efficiency with which the magnetic energy is converted into X-ray radiation, and $A$ represents the effective area over which the Poynting flux is injected into the corona. In our cases, the $A$ is the same ($A = 4\pi R_\odot^2$). If $\eta_{\rm rad}$ is approximately constant, this leads to $L_X \propto B^{1.675}$. The slight mismatch between the fitted exponent (1.75) and the theoretical one (1.675) likely stems from local variations in plasma conditions (e.g., density, temperature) and multiple energy-dissipation channels, which break the strict linear conversion of Poynting flux into X-ray output, and such 0.1–0.2 deviations are fairly common in complex MHD models. The fraction of the energy emitted within a given bandpass (e.g., 0.1–2.4 keV in our analysis) depends sensitively on the coronal EMDs and the radiation mechanisms (J. Sanz-Forcada et al. 2011). All these effects can contribute to the deviation of the fitted power-law index from the theoretical value.

Given that our Alfvén wave-driven model, based on an extended solar coronal framework, successfully reproduces key observational trends across solar-type stars, our results suggest that Alfvén wave heating could be a viable and broadly applicable mechanism in solar-type stars. Nonetheless, we note that the present model does not include other potential coronal heating mechanisms, such as nanoflares or additional wave-based processes, which may contribute to energy deposition in the corona (E. N. Parker 1988; J. A. Klimchuk 2006; S. R. Cranmer & A. R. Winebarger 2019; T. Van Doorsselaere et al. 2025). Investigating the influence of these mechanisms in global MHD models remains a significant challenge for future studies.

### 4.2. Implications for Eruptive Events

Modeling quiescent coronae provides a foundation for further modeling eruptive events. As demonstrated by the coronal structures in our models, the magnetic fields of rapidly rotating solar-type stars can be tens to hundreds of times stronger than that of the Sun, whereas the plasma density and temperature increase by only about 1 order of magnitude. As a result, the coronae exhibit significantly higher Alfvén velocities ($V_A \approx 10^3$–$10^4$ km s$^{-1}$, where $V_A = B/\sqrt{\mu\rho}$, within about 2 $R_\star$) and the extended low-plasma-beta regions ($\beta \ll 1$, where $\beta = nkT/\frac{B^2}{\mu_0}$, indicating the magnetic-field-dominated regime). In the lower corona, plasma can thus be accelerated above 1000 km s$^{-1}$ more easily, likely altering the generation and early formation conditions of coronal mass ejections and other eruptive phenomena. Another feature is that solar-like eruptions may occur in a more strongly constrained and large-scale magnetic environment that easily leads to failed eruptions (J. D. Alvarado-Gómez et al. 2018, 2019, 2020).

Moreover, eruptive events may display spatial distributions and timescales distinct from those in the solar corona. For instance, in a magnetic-field-dominated environment, flares may form larger post-flare loops, resulting in prolonged cooling timescales and higher total radiative output (M. J. Aschwanden et al. 2008; H. Maehara et al. 2012). Meanwhile, the plasma might be easily heated to 100 MK, which will further enhance high-energy emission and particle acceleration, thus observed as superflares (A. O. Benz 2017; A. F. Kowalski 2024).

## 5. Conclusions

In this work, we employed the SWMF-AWSoM model to simulate solar and stellar atmospheres across a range of stellar rotation rates and magnetic activity levels. Instead of using ZDI-based reconstructions, we employed dynamo-generated magnetic maps for four solar-type stars ($M_\star = 1\,M_\odot$, $R_\star = 1\,R_\odot$) with different rotation rates, sampling both magnetic minimum and maximum conditions. Two solar magnetograms from SDO/HMI were used for comparison. The stellar rotation rates in our model thus span from $1\,\Omega_\odot$ (corresponding to a period of 25.38 days) to $23.3\,\Omega_\odot$ (1.09 days). It is known that with increasing rotation rate, the average magnetic field strength $\langle|\boldsymbol{B}|\rangle$ also increases (for unsaturated stars). For the Sun, the mean magnetic field strengths are 6.02 G (Sun$_{\rm min}$) and 27.7 G (Sun$_{\rm max}$), while for solar-type stars, we scaled the values in the range of 316–1210 G.

The inclusion of small-scale magnetic field information enables us to analyze the detailed structure of stellar coronae and provides a sufficient energy budget to heat the corona. The typical electron number densities at a radial distance of 1.5 $R_\star$ are 2–8 $\times 10^8$ cm$^{-3}$, which are 1–2 orders of magnitude higher than solar values (in our model, it is $\sim 10^7$ cm$^{-3}$). The closed magnetic field regions confine dense and ultra-hot plasma. In the solar corona, the peak temperatures reach 1–4 $\times 10^6$ K, while in solar-type stars, the peak coronal temperatures can reach $1 \times 10^7$ K. These hot regions extend far beyond the stellar surface, reaching distances of several stellar radii.

We compared the EMDs from our models with those derived from observations of active stars and found that our models successfully reproduce the hot ($10^{6.2} < T < 10^{6.8}$ K) and ultra-hot ($T > 10^{6.8}$ K) plasma components responsible for most of the X-ray emission. The cooler plasma ($T < 10^{6.2}$ K) is systematically overestimated due to limitations of the model. This affects the accuracy of our calculations in the EUV range, but does not impact the conclusions about the soft X-ray band. Furthermore, we used the CHIANTI atomic database to compute synthetic stellar spectra. The resulting spectra for solar-type stars exhibit prominent emission lines, including Si XIII and Si XIV (at ∼1.9 keV), Ca XVIII and Ca XIX (∼3.9 keV), and Fe XX–XXV (∼6.6 and 7.8 keV). These lines are formed at $\log T \approx 7.0$–7.1 and can serve as useful diagnostics of stellar coronal conditions in future missions. We used the synthetic X-ray spectra to derive the X-ray luminosities ($L_X$, ROSAT PSPC 0.1–2.4 keV band luminosities) of our models. The resulting values closely match the observed relation between $L_X/L_{\rm bol}$ and the Rossby number, providing further validation of our models. Additionally, we derived a scaling law $L_X \propto \langle|\boldsymbol{B}|\rangle^{1.75}$, which suggests a reduced efficiency in converting magnetic energy ($\propto B^2$) into X-ray emission at higher field strengths.

In synthetic X-ray images, bright X-ray emissions are from the high-temperature, high-density regions associated with closed magnetic field lines. In contrast, the dark coronal holes are related to open field regions that are less dense and cooler. The spatially inhomogeneous distribution of bright active regions and dark coronal holes leads to significant rotational modulation of the X-ray emission.

This study demonstrates that incorporating small-scale magnetic field structures from global convective dynamo (GCD) models enables the reproduction of quiescent coronae in solar-type stars. Unlike earlier studies that estimate coronal properties using empirical scaling laws or simplified loop models, our approach solves the full set of MHD equations starting from physically generated surface magnetic fields. This yields a self-consistent, three-dimensional coronal structure, capturing not only the plasma spatial distribution and X-ray morphology, but also their physical relation with magnetic field strength and stellar rotation. These results provide a crucial foundation for extending current models toward eruptive phenomena. Future work should aim to couple steady-state coronal models with time-dependent simulations of eruptive events, explore a wider range of stellar magnetic topologies, and incorporate observational constraints from next-generation multiwavelength instruments. Such efforts will be key to understanding the magnetic activity of solar analogs and its broader implications for stellar evolution and planetary habitability.

## Acknowledgments

The numerical calculations in this paper were done on the computing facilities in the High Performance Computing Center of Nanjing University. SDO is a mission of NASA's Living With a Star (LWS) program. Hinode is a Japanese mission developed and launched by ISAS/JAXA, with NAOJ as a domestic partner and NASA and STFC (UK) as international partners. It is operated by these agencies in cooperation with ESA and NSC (Norway). CHIANTI is a collaborative project involving George Mason University, the University of Michigan (USA), University of Cambridge (UK), and NASA Goddard Space Flight Center (USA). This work is supported by the National Natural Science Foundation of China under grants 12525305, 12403066, the Fundamental Research Funds for the Central Universities under grants 2024300348 and 2025300318, the Jiangsu Funding Program for Excellent Postdoctoral Talent, and the Postgraduate Research & Practice Innovation Program of Jiangsu Province (Project No. KYCX24_0182).

## Appendix A
## Methods of Spectrum Calculations

In this work, the continuum and spectral line intensities were calculated using the CHIANTI Spectral Synthesis Package (v11.0; K. P. Dere et al. 1997; R. P. Dufresne et al. 2024), which assumes ionization equilibrium and utilizes atomic data optimized for high-temperature, optically thin plasmas.

The input electron temperature ($T_e$) and density ($n$) profiles were obtained from model data. The differential emission measure (DEM) was specified in a volume-integrated form as $\mathrm{DEM}(T_i) = \frac{\mathrm{EM}(T_i)}{\Delta T_i}$, rather than as a line-of-sight column DEM. Consequently, the CHIANTI output is considered to already incorporate the emitting area.

We used a uniform temperature bin of $\Delta \log T = 0.1$ for the temperature range of $6.2 \leqslant \log T \leqslant 8.0$, resulting in $N = 18$. Note that $\Delta T_i$ is not constant, as the temperature axis is logarithmically spaced; accordingly, $\Delta T_i$ increases with temperature.

The emergent spectral intensity was computed using the following expression:

$$I(\lambda) = \sum_i \left[\varepsilon_{\mathrm{cont}}(\lambda, T_i) + \sum_j \varepsilon_{\mathrm{line},j}(\lambda, T_i)\right] \mathrm{DEM}(T_i)\Delta T_i,$$

where $\varepsilon_{\mathrm{cont}}(\lambda, T)$ represents the continuum emissivity, primarily dominated by free–free emission, with additional contributions from free–bound processes and two-photon emission. The line emissivity $\varepsilon_{\mathrm{line},j}(\lambda, T)$ refers to the contribution of the $j$th spectral line, which depends directly on the elemental abundances.

Elemental abundances were set using the `Sun_coronal_2021_chianti.abund` file from the `SSWIDL` distribution, which enhances low first ionization potential (FIP) elements by 0.5 dex relative to the photospheric abundances of M. Asplund et al. (2021). This corresponds to an FIP bias of 3.16, consistent with typical coronal values reported for active regions (U. Feldman et al. 1992).

The calculation was performed over a wavelength range of 1.24–124 Å (0.1–10 keV), and the resulting spectrum was subsequently converted to photon energy space (keV) to produce an energy-resolved spectrum.

## Appendix B
## Methods of X-Ray Imaging

We used the `xrt_flux1000.pro` routine from `SSWIDL` to calculate the response $G(T)$ of Hinode/XRT thick-Be filter over the range $\log T = 5.0$–$8.0$ (Figure 7). The filter is primarily sensitive to high-temperature plasma, with minimal contribution from material cooler than $\log T < 6.2$.

Synthetic XRT images were generated using a built-in line-of-sight integration routine from the SWMF framework, which produces as $384 \times 384$ image per viewing angle. A circular mask with a radius of $3\,R_\star$ was applied to define the region of interest. For each pixel ($x$, $y$), the algorithm computes the column differential emission measure (CDEM), defined as:

$$\mathrm{CDEM}(T_i, x, y) = \frac{1}{\Delta T_i} \int_{z}^{} {}_{T - \frac{\Delta T_i}{2} \leqslant T < T + \frac{\Delta T_i}{2}} n^2 \mathrm{d}z,$$

where $n(x, y, z)$ is the electron number density in units of $\mathrm{cm}^{-3}$, and the resulting CDEM has units of $\mathrm{cm}^{-5}$ K.

The corresponding XRT intensity at each pixel is then computed by convolving CDEM with the $G(T)$ using bins spaced by $\Delta \log T = 0.1$:

$$I_{\mathrm{XRT}}(x, y) = \sum_i \mathrm{CDEM}(T_i, x, y) G(T_i)\Delta T_i.$$

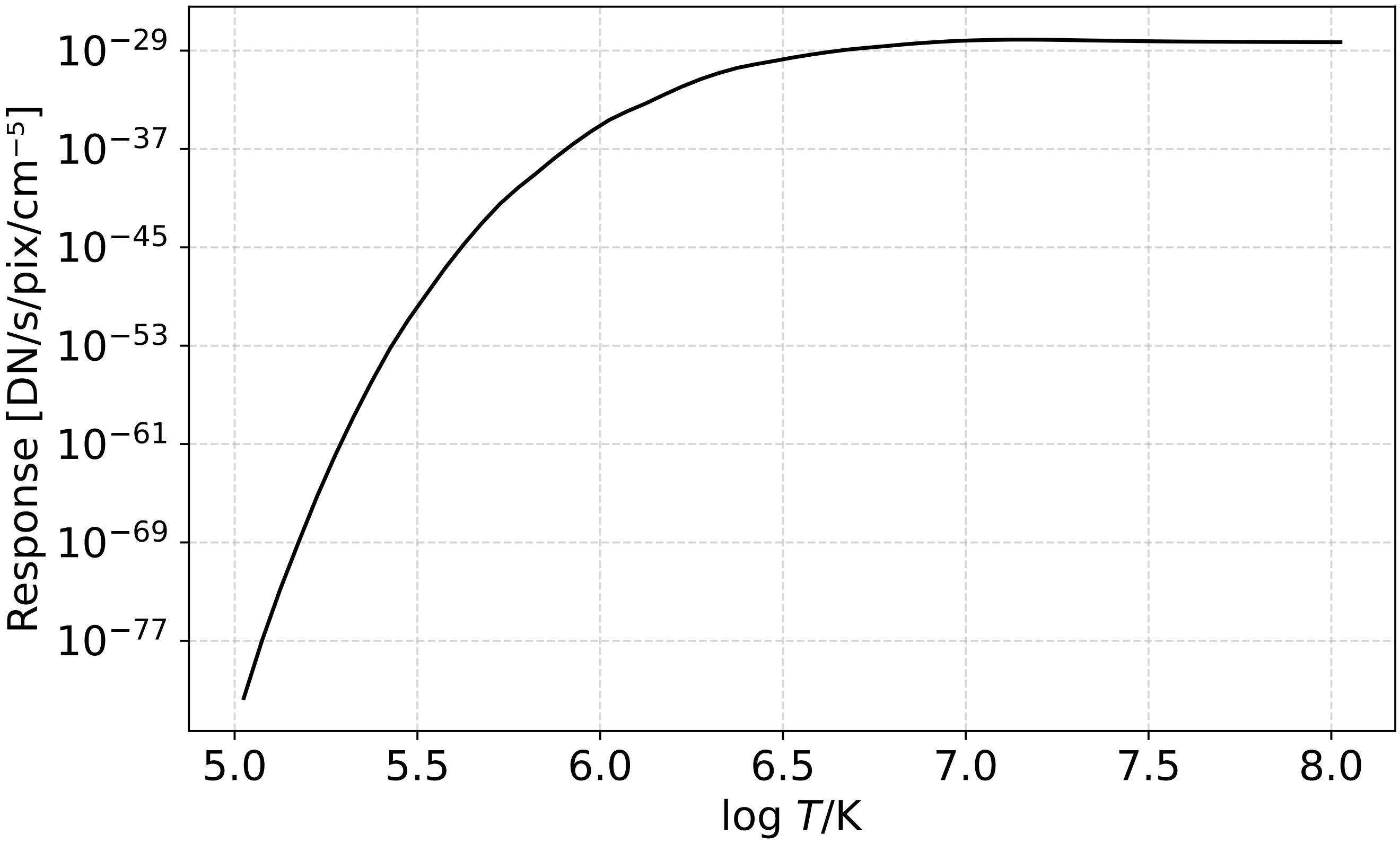

**Figure 7.** Temperature response function of the thick beryllium (thick-Be) filter from Hinode/XRT.

## ORCID iDs

Yue-Hong Chen (陈悦虹) https://orcid.org/0000-0003-1220-1582
Julián D. Alvarado-Gómez https://orcid.org/0000-0001-5052-3473
Xin Cheng (程鑫) https://orcid.org/0000-0003-2837-7136
Yu Dai (戴煜) https://orcid.org/0000-0001-9856-2770
Tong Shi https://orcid.org/0000-0002-0827-7632
Katja Poppenhäger https://orcid.org/0000-0003-1231-2194
Chen Xing (邢晨) https://orcid.org/0000-0002-6550-1522
Shun Inoue https://orcid.org/0000-0003-3085-304X
Jörn Warnecke https://orcid.org/0000-0002-9292-4600
Maarit J. Korpi-Lagg https://orcid.org/0000-0002-9614-2200
Mingde Ding (丁明德) https://orcid.org/0000-0002-4978-4972

## References

Alvarado-Gómez, J. D., Drake, J. J., Cohen, O., Moschou, S. P., & Garraffo, C. 2018, ApJ, 862, 93
Alvarado-Gómez, J. D., Drake, J. J., Fraschetti, F., et al. 2020, ApJ, 895, 47
Alvarado-Gómez, J. D., Drake, J. J., Moschou, S. P., et al. 2019, ApJL, 884, L13
Alvarado-Gómez, J. D., Hussain, G. A. J., Cohen, O., et al. 2016a, A&A, 588, A28
Alvarado-Gómez, J. D., Hussain, G. A. J., Cohen, O., et al. 2016b, A&A, 594, A95
Antiochos, S. K. 1980, ApJ, 241, 385
Aschwanden, M. J., & Benz, A. O. 1997, ApJ, 480, 825
Aschwanden, M. J., Stern, R. A., & Güdel, M. 2008, ApJ, 672, 659
Asplund, M., Amarsi, A. M., & Grevesse, N. 2021, A&A, 653, A141
Barnes, S. A. 2003, ApJ, 586, 464
Benz, A. O. 2017, LRSP, 14, 2
Boro Saikia, S., Lueftinger, T., Airapetian, V. S., et al. 2023, ApJ, 950, 124
Brickhouse, N. S., & Dupree, A. K. 1998, ApJ, 502, 918
Brun, A. S., & Browning, M. K. 2017, LRSP, 14, 4
Chandra, S., Vats, H. O., & Iyer, K. N. 2010, MNRAS, 407, 1108
Chandran, B. D. G., Dennis, T. J., Quataert, E., & Bale, S. D. 2011, ApJ, 743, 197
Charbonneau, P. 2014, ARA&A, 52, 251
Charbonneau, P., & Sokoloff, D. 2023, SSRv, 219, 35
Chebly, J. J., Alvarado-Gómez, J. D., Poppenhäger, K., & Garraffo, C. 2023, MNRAS, 524, 5060
Coffaro, M., Stelzer, B., Orlando, S., et al. 2020, A&A, 636, A49
Cranmer, S. R., & Saar, S. H. 2011, ApJ, 741, 54
Cranmer, S. R., & Winebarger, A. R. 2019, ARA&A, 57, 157
Dere, K. P., Landi, E., Mason, H. E., Monsignori Fossi, B. C., & Young, P. R. 1997, A&AS, 125, 149
Donati, J. F., Cristofari, P. I., Klein, B., Finociety, B., & Moutou, C. 2025, A&A, 700, A122
Donati, J. F., Semel, M., Carter, B. D., Rees, D. E., & Collier Cameron, A. 1997, MNRAS, 291, 658
do Nascimento, J. D., Jr., Vidotto, A. A., Petit, P., et al. 2016, ApJL, 820, L15
Drake, J. J., & Stelzer, B. 2023, Handbook of X-ray and Gamma-ray Astrophysics (Berlin: Springer), 132
Dufresne, R. P., Del Zanna, G., Young, P. R., et al. 2024, ApJ, 974, 71
Evensberget, D., Carter, B. D., Marsden, S. C., Brookshaw, L., & Folsom, C. P. 2021, MNRAS, 506, 2309
Evensberget, D., Carter, B. D., Marsden, S. C., et al. 2022, MNRAS, 510, 5226
Evensberget, D., Marsden, S. C., Carter, B. D., et al. 2023, MNRAS, 524, 2042
Feldman, U., Mandelbaum, P., Seely, J. F., Doschek, G. A., & Gursky, H. 1992, ApJS, 81, 387
Garraffo, C., Drake, J. J., & Cohen, O. 2015, ApJL, 807, L6
Golub, L., DeLuca, E., Austin, G., et al. 2007, SoPh, 243, 63
Gombosi, T. I., Chen, Y., Glocer, A., et al. 2021, JSWSC, 11, 42
Gombosi, T. I., Powell, K. G., De Zeeuw, D. L., et al. 2004, CSE, 6, 14
Gronoff, G., Arras, P., Baraka, S., et al. 2020, JGRA, 125, e27639
Güdel, M. 2004, A&ARv, 12, 71
Güdel, M., Audard, M., Kashyap, V. L., Drake, J. J., & Guinan, E. F. 2003, ApJ, 582, 423
Güdel, M., & Nazé, Y. 2009, A&ARv, 17, 309
Hackman, T., Kochukhov, O., Viviani, M., et al. 2024, A&A, 682, A156
Hollweg, J. V. 1978, RvGSP, 16, 689
Huang, Z., Tóth, G., Sachdeva, N., et al. 2023, ApJL, 946, L47
Inoue, S., Iwakiri, W. B., Enoto, T., et al. 2024, ApJL, 969, L12

Jian, L. K., MacNeice, P. J., Mays, M. L., et al. 2016, SpWea, 14, 592
Jin, M., Manchester, W. B., van der Holst, B., et al. 2017a, ApJ, 834, 172
Jin, M., Manchester, W. B., van der Holst, B., et al. 2017b, ApJ, 834, 173
Judge, P. G., Solomon, S. C., & Ayres, T. R. 2003, ApJ, 593, 534
Käpylä, P. J., Browning, M. K., Brun, A. S., Guerrero, G., & Warnecke, J. 2023, SSRv, 219, 58
Klimchuk, J. A. 2006, SoPh, 234, 41
Kochukhov, O., Hackman, T., Lehtinen, J. J., & Wehrhahn, A. 2020, A&A, 635, A142
Kowalski, A. F. 2024, LRSP, 21, 1
Kurihara, M., Tsujimoto, M., Audard, M., et al. 2025, PASJ, S77, 50
Lionello, R., Linker, J. A., & Mikić, Z. 2009, ApJ, 690, 902
Lithwick, Y., Goldreich, P., & Sridhar, S. 2007, ApJ, 655, 269
Maehara, H., Shibayama, T., Notsu, S., et al. 2012, Natur, 485, 478
Maggio, A., Pillitteri, I., Argiroffi, C., et al. 2024, A&A, 690, A383
Manchester, W. B., IV, van der Holst, B., & Lavraud, B. 2014, PPCF, 56, 064006
Manchester, W. B., IV, van der Holst, B., Tóth, G., & Gombosi, T. I. 2012, ApJ, 756, 81
Matthaeus, W. H., Zank, G. P., Oughton, S., Mullan, D. J., & Dmitruk, P. 1999, ApJL, 523, L93
Meng, X., van der Holst, B., Tóth, G., & Gombosi, T. I. 2015, MNRAS, 454, 3697
Morin, J., Donati, J. F., Petit, P., et al. 2008, MNRAS, 390, 567
Nicholson, B. A., Vidotto, A. A., Mengel, M., et al. 2016, MNRAS, 459, 1907
Ó Fionnagáin, D., Vidotto, A. A., Petit, P., et al. 2019, MNRAS, 483, 873
Oran, R., van der Holst, B., Landi, E., et al. 2013, ApJ, 778, 176
Owen, J. E., & Wu, Y. 2013, ApJ, 775, 105
Pallavicini, R., Golub, L., Rosner, R., et al. 1981, ApJ, 248, 279
Parker, E. N. 1988, ApJ, 330, 474
Pesnell, W. D., Thompson, B. J., & Chamberlin, P. C. 2012, SoPh, 275, 3
Pevtsov, A. A., Fisher, G. H., Acton, L. W., et al. 2003, ApJ, 598, 1387
Pognan, Q., Garraffo, C., Cohen, O., & Drake, J. J. 2018, ApJ, 856, 53
Powell, K. G., Roe, P. L., Linde, T. J., Gombosi, T. I., & De Zeeuw, D. L. 1999, JCoPh, 154, 284
Prša, A., Harmanec, P., Torres, G., et al. 2016, AJ, 152, 41
Reale, F. 2007, A&A, 471, 271
Reiners, A., Shulyak, D., Käpylä, P. J., et al. 2022, A&A, 662, A41
Rempel, M., Bhatia, T., Bellot Rubio, L., & Korpi-Lagg, M. J. 2023, SSRv, 219, 36
Réville, V., Folsom, C. P., Strugarek, A., & Brun, A. S. 2016, ApJ, 832, 145
Ribas, I., Guinan, E. F., Güdel, M., & Audard, M. 2005, ApJ, 622, 680
Rosner, R., Tucker, W. H., & Vaiana, G. S. 1978, ApJ, 220, 643
Sachdeva, N., Manchester, W. B., IV, Sokolov, I., et al. 2023, ApJ, 952, 117
Sachdeva, N., Tóth, G., Manchester, W. B., et al. 2021, ApJ, 923, 176
Sachdeva, N., van der Holst, B., Manchester, W. B., et al. 2019, ApJ, 887, 83
Sanz-Forcada, J., López-Puertas, M., Lampón, M., et al. 2025, A&A, 693, A285
Sanz-Forcada, J., Micela, G., Ribas, I., et al. 2011, A&A, 532, A6
Sanz-Forcada, J., Stelzer, B., Coffaro, M., Raetz, S., & Alvarado-Gómez, J. D. 2019, A&A, 631, A45
Sasaki, R., Tsuboi, Y., Iwakiri, W., et al. 2021, ApJ, 910, 25
Schou, J., Scherrer, P. H., Bush, R. i., et al. 2012, SoPh, 275, 229
Schwenn, R. 2006, LRSP, 3, 2
Semel, M. 1989, A&A, 225, 456
Shi, T., Manchester, W., IV, Landi, E., et al. 2022, ApJ, 928, 34
Shi, T., Manchester, W., Landi, E., et al. 2024, ApJ, 961, 60
Sokolov, I. V., van der Holst, B., Oran, R., et al. 2013, ApJ, 764, 23
Spitzer, L., & Härm, R. 1953, PhRv, 89, 977
Tanaka, K., Watanabe, T., & Nitta, N. 1984, ApJ, 282, 793
Tóth, G., Sokolov, I. V., Gombosi, T. I., et al. 2005, JGRA, 110, A12226
Tóth, G., van der Holst, B., Sokolov, I. V., et al. 2012, JCoPh, 231, 870
van der Holst, B., Huang, J., Sachdeva, N., et al. 2022, ApJ, 925, 146
van der Holst, B., Manchester, W. B., IV, Frazin, R. A., et al. 2010, ApJ, 725, 1373
van der Holst, B., Sokolov, I. V., Meng, X., et al. 2014, ApJ, 782, 81
Van Doorsselaere, T., Sieyra, M. V., Magyar, N., Goossens, M., & Banović, L. 2025, A&A, 696, A166
Velli, M., Grappin, R., & Mangeney, A. 1989, PhRvL, 63, 1807
Vidotto, A. A. 2021, LRSP, 18, 3
Vidotto, A. A., Fares, R., Jardine, M., Moutou, C., & Donati, J. F. 2015, MNRAS, 449, 4117
Vidotto, A. A., Gregory, S. G., Jardine, M., et al. 2014, MNRAS, 441, 2361
Viviani, M., Warnecke, J., Käpylä, M. J., et al. 2018, A&A, 616, A160
Vogt, S. S., Penrod, G. D., & Hatzes, A. P. 1987, ApJ, 321, 496
Wargelin, B. J., Saar, S. H., Irving, Z. A., et al. 2024, ApJ, 977, 144
Wood, B. E. 2018, JPhCS, 1100, 012028
Wright, N. J., Drake, J. J., Mamajek, E. E., & Henry, G. W. 2011, ApJ, 743, 48
Wright, N. J., Newton, E. R., Williams, P. K. G., Drake, J. J., & Yadav, R. K. 2018, MNRAS, 479, 2351
Xu, Y., Tian, H., Alvarado-Gómez, J. D., Drake, J. J., & Guerrero, G. 2025, ApJ, 985, 219
Zank, G. P., Dosch, A., Hunana, P., et al. 2012, ApJ, 745, 35
Zank, G. P., Matthaeus, W. H., & Smith, C. W. 1996, JGR, 101, 17093